\documentclass{aastex62}
\usepackage{amsmath}
\usepackage{amssymb}
\usepackage{graphicx}

\begin{document}
\title{The $H_0$ Tension in Non-flat QCDM Cosmology}
\author{Haitao Miao}
\affiliation{School of Physics and Astronomy, Sun Yat-sen University, 2 Daxue Rd, Tangjia, Zhuhai, China}
\author{Zhiqi Huang}
\affiliation{School of Physics and Astronomy, Sun Yat-sen University, 2 Daxue Rd, Tangjia, Zhuhai, China}

\correspondingauthor{Zhiqi Huang}

\begin{abstract}
  The recent local measurement of Hubble constant leads to a more than $3\sigma$ tension with Planck + $\Lambda$CDM \citep{Riess18}. In this article we study the $H_0$ tension in non-flat QCDM cosmology, where Q stands for a minimally coupled and slowly-or-moderately rolling quintessence field $\phi$ with a smooth potential $V(\phi)$. By generalizing the QCDM one-parameter and three-parameter parametrizations in \citet{HBK} to non-flat universe and using the latest cosmological data, we find that the $H_0$ tension remains above $3.2\sigma$ level for this class of model.

\end{abstract}


\section{Introduction}\label{sec:intro}

Confirmed late-time acceleration of the Universe by observational data including Type Ia supernovae (SNe) \citep{Riess98, Perlmutter99, JLA}, cosmic microwave background (CMB) radiation  \citep{PlanckOverview15, PlanckParam15, PlanckLike15, PlanckDE15} and baryon acoustic oscillations (BAO) \citep{BAO-SDSS-6DF, BAO-SDSS-DR7-MGS, BAO-SDSS-DR12-CMASS, BAO-SDSS-DR12-LOWZ} indicates that about 70\% of the energy density of the Universe today consists of dark energy, which is supposed exotic component of the Universe that induces a negative pressure on large scales.

In the standard $\Lambda$CDM cosmology, the late-time cosmic acceleration is explained by Einstein's cosmological constant $\Lambda$, whose microscopic nature is interpreted as vacuum energy. There is, however, a serious fine-tuning problem with this interpretation: the measured energy scale of $\Lambda$ is $\sim 10^{120}$ times smaller than a naive dimension analysis. This ``discrepancy'' motivated theorists to construct alternative models, among which the first suggestion is quintessence, namely, a minimally coupled canonical scalar field with a potential $V(\phi)$ \citep{Wetterich88, Ratra88, Caldwell98, Zlatev99}. A slowly rolling quintessence field can provide a negative pressure that drives the cosmic acceleration. Physicists have also proposed many other ``more exotic'' models, such as k-essence \citep{Armendariz-Picon00, Armendariz-Picon01}, $f(R)$ gravity \citep{Capozziello03, Carroll04, Nojiri06, Hu07}, and DGP model \citep{Dvali00}. For a comprehensive list of dark energy models, the reader is referred to \cite{Copeland06}, \cite{Yoo12} and \cite{Arun17}.

Over the last two decades, $\Lambda$CDM model continued to be the simplest model to explain the observational data,  and other candidates seemed to be disfavored by Occam's razor. Recently, the improved local measurement of the Hubble constant $H_0$ starts to challenge this picture: the locally measured $H_0 = 73.48 \pm 1.66\, \mathrm{km/s/Mpc}$ \citep{Riess18} is about $3\sigma$ (or even more, depending on which combination of CMB data sets is used) higher than the CMB + $\Lambda$CDM favored value $H_0=67.8\pm 0.9\,  \mathrm{km/s/Mpc}$ \citep{PlanckParam15}.

In minimally coupled or weakly coupled dark energy models the dark energy component can often be approximated as a perfect fluid. The equation of state (EOS) of the fluid, defined as the ratio of the pressure to the energy density, depending on the underlying model can be approximately a constant or strongly time-dependent. For $\Lambda$CDM model, the EOS is equal to $-1$. Whereas for a quintessence field, the EOS is characterized by a time-dependent function $w(a) \ge -1$, $a$ being the scale factor in the Friedmann-Robertson-Walker (FRW) metric. To make definitive predictions that can be compared with the data, observers need to specify a function form $w(a)$. Chevallier-Polarski-Linder (CPL) \citep{Chevallier01,Linder03} parametrization, $w=w_0+(1-a)w_a$, is the most popular one in the literature. However, such a parametrization is not based on any physical model. This arbitrariness in CPL parametrization leads people to investigate more theoretically motivated dark energy trajectories described by a few physical parameters. For quintessence models, an analytic approximation of $w(a)$ has been derived in \cite{HBK} (HBK). HBK's three-parameter approximation $w(a; \varepsilon_s, \varepsilon_{\phi\infty}, \zeta_s)$ fits well the ensemble of trajectories for a wide class of potentials $V(\phi)$. The slope parameter $\varepsilon_s$ characterizes the slope of the potential. They found that a reasonable pivot to measure the slope of the potential is at $a=a_{\rm eq}$, where the energy densities of dark energy and matter are equal, and so is $\varepsilon_s$ defined.  The tracking parameter $\varepsilon_{\phi\infty}$ and the running parameter $\zeta_s$ induce necessary corrections if the quintessence has early-time dynamics.

HBK's $w$ formula was based on flat-space assumption, a well established observational fact in $\Lambda$CDM cosmology. In other words, HBK implicitly assumed that the constraint on the spatial curvature is not sensitive to the choice of dark energy model. In this article we re-examine this assumption by generalizing HBK parametrization to FRW metric with non-vanishing spatial curvature.

This article is organized as follows. Section \ref{sec:parametrization} generalize the HBK $w$ parametrization to non-flat FRW metric; Section \ref{sec:data} describes the observational data sets. In Section \ref{sec:results} we constrain the generalized HBK $w$ parametrization and study the $H_0$ tension in this class of model; Section \ref{sec:conclusion} concludes.

Throughout the article we work with natural units with $c=\hbar=1$.  The reduced Planck mass is defined as $M_p \equiv \frac{1}{\sqrt{8\pi G_N}}$, where $G_N$ is Newton's gravitational constant. We assume three species of light neutrinos with default sum of mass $\sum m_\nu = 0.06 \mathrm{eV}$.

\section{Generalizing the HBK Parametrization \label{sec:parametrization}}

We begin with FRW metric
\begin{equation}
ds^2 = dt^2-a^2(t)\left[\frac{1}{1-kr^2} dr^2 + r^2 \left(d\theta^2+\sin\theta^2d\phi^2\right)\right]. \label{eq:FRW}
\end{equation}
The scale factor $a$ is normalized to unity today and can be related to cosmological redshift $z$ via $a=\frac{1}{1+z}$. Cosmological expansion is characterized by the Hubble parameter
\begin{equation}
  H \equiv \frac{\dot a}{a},
\end{equation}
where a dot denotes time derivative $d/dt$.

The Hubble parameter today, namely the Hubble constant is denoted as $H_0$. The combination of $H_0$ and the parameter $k$ in FRW metric yields a dimensionless parameter
\begin{equation}
  \Omega_k \equiv -\frac{k}{H_0^2},
\end{equation}
which characterizes the spatial curvature of the current universe.

At the background level, the Klein-Gordon equation of the quintessence field is
\begin{equation}
  \ddot \phi + 3H\dot\phi + V'(\phi) = 0, \label{eq:eom}
\end{equation}
where $V'(\phi)=dV/d\phi$.

The pressure of quintessence field is the difference between the kinetic energy and the potential energy:
\begin{equation}
  p_\phi = \frac{1}{2}\dot\phi^2-V(\phi).
\end{equation}
The energy density is the sum:
\begin{equation}
  \rho_\phi = \frac{1}{2}\dot\phi^2+V(\phi).
\end{equation}
Because a slowly rolling quintessence field has a sub-dominant kinetic energy $\frac{1}{2}\dot\phi^2 \ll V(\phi)$, its EOS $w=p_\phi/\rho_\phi$ is close to $-1$. To the lowest order approximation, dark energy behaves like a cosmological constant: $1+w\approx  0$ and $\rho_\phi \approx const$. Our goal is to compute the next order correction $1+w$, or equivalently, the time evolution of $\rho_\phi$.

It is useful to introduce a dimensionless parameter
\begin{equation}
  \theta \equiv  \arcsin{\frac{\dot\phi}{\sqrt{2\rho_\phi}}}.
\end{equation}
Using equation \eqref{eq:eom} and after a few lines of algebra,  we achieve
\begin{equation}
  \frac{d\theta}{d\ln a}= \sqrt{\epsilon_V}\frac{\sqrt{\rho_\phi}}{HM_p}\cos\theta-\frac{3}{2}\sin{2\theta},
\end{equation}
where
\begin{equation}
  \epsilon_V \equiv \frac{M_p^2}{2}\left(\frac{V'}{V}\right)^2
\end{equation}
describes the slope of the (logarithm) potential.

At low redshift where dark energy is relevant, we can ignore the radiation and light neutrinos and compute the Hubble parameter via
\begin{equation}
  H = H_0\sqrt{\Omega_m a^{-3} +\Omega_k a^{-2}+  \frac{\rho_\phi}{3H_0^2M_p^2} },
\end{equation}
where $\Omega_m$ is the ratio of today's matter density to the critical density $\rho_{\rm crit} = 3H_0^2M_p^2$. Similarly we denote today's value of $ \frac{\rho_\phi}{3H_0^2M_p^2}$ as $\Omega_\phi$.

We proceed with slow-roll approximation $\theta\ll 1$. Following \cite{HBK}, we bootstrap from the zeroth order approximations $\rho_\phi \approx 3H_0^2M_p^2\Omega_\phi$ and $\epsilon_V \approx \varepsilon_s$. The evolution equation of $\theta$ approximated to the lowest order reads
\begin{equation}
  \frac{d\theta}{d \ln a } \approx \sqrt{\frac{3\varepsilon_s\Omega_\phi}{\Omega_m a^{-3} + \Omega_k a^{-2} + \Omega_\phi}}- 3\theta. \label{eq:approx}
\end{equation}
For the moment we assume no early-time dynamics of the quintessence field, i.e.,
\begin{equation}
  \theta_{a\rightarrow 0^+}  = 0. \label{eq:ic}
\end{equation}
The solution for \eqref{eq:approx} and \eqref{eq:ic} is
\begin{equation}
  \theta \approx \sqrt{\frac{\varepsilon_s}{3}}\,F\left(\frac{\Omega_k}{\Omega_m}\left(\frac{\Omega_m}{\Omega_\phi}\right)^{\frac{1}{3}}, a\left(\frac{\Omega_\phi}{\Omega_m}\right)^{\frac{1}{3}}\right),
\end{equation}
where
\begin{equation}
  F(\lambda, x) \equiv \frac{3}{x^3} \int_0^x \sqrt{\frac{t^7}{1+\lambda t+t^3}}
  \,dt .
\end{equation}
In HBK's parametrization, $\lambda = 0$ and
\begin{equation}
  F(0, x) = \frac{\sqrt{1+x^3}}{x^{3/2}}-\frac{\ln\left[x^{3/2}+\sqrt{1+x^3}\right]}{x^3} \, 
\end{equation}
is instead used.

The EOS is then
\begin{equation}
  w \approx -1 + 2\theta^2 \approx -1 + \frac{2\varepsilon_s}{3} \,F^2\left(\frac{\Omega_k}{\Omega_m}\left(\frac{\Omega_m}{\Omega_\phi}\right)^{\frac{1}{3}}, a\left(\frac{\Omega_\phi}{\Omega_m}\right)^{\frac{1}{3}}\right). \label{eq:w}
\end{equation}
A seemingly trivial but key observation is that the above formula can be achieved by replacing $F(0, x)$ in HBK's one-parameter approximation with the full $F(\lambda, x)$. 

Eq.~\eqref{eq:w} as a generalization of HBK's one-parameter parametrization is valid for flat-potential models where the scalar field is ``frozen'' ($\dot\phi\rightarrow 0$ and $w\rightarrow -1$) by a large Hubble friction in the early Universe. To cover another popular class of models where the scalar field has a ``tracking'' behavior ($w\sim \mathrm{constant}> -1$) in the early Universe,  HBK proposed a three-parameter approximation $w(a; \varepsilon_s, \varepsilon_{\phi\infty}, \zeta_s)$. The ``slope parameter'' $\varepsilon_s$ is defined as $\epsilon_V$ at ``matter-DE density equality'' $a=a_{\rm eq}$.  The ``tracking parameter'' $\varepsilon_{\phi\infty}$ is defined as $\frac{\epsilon_V\rho_\phi}{3H^2M_p^2}$ at $a\ll a_{\rm eq}$. (For tracking models this quantity is approximately a constant.) The ``running parameter'' is defined in a more sophisticated way
\begin{equation}
\zeta_s \equiv\frac{ \left.\frac{dq}{dy}\right\vert_{a=a_{\rm eq}} - \left.\frac{dq}{dy}\right\vert_{a\rightarrow 0^+}}{\left.\frac{dq}{dy}\right\vert_{a=a_{\rm eq}} + \left.\frac{dq}{dy}\right\vert_{a\rightarrow 0^+}} , \label{parrundef}
\end{equation}
where
\begin{equation}
q \equiv \frac{\sqrt{\epsilon_V\rho_{\phi}}}{H},\ y\equiv \sqrt{ \frac{\left(\frac{a}{a_{\rm eq}}\right)^3}{1+ \left(\frac{a}{a_{\rm eq}} \right)^3}} \,\label{eq:y}. 
\end{equation}
HBK showed that $\zeta_s$ is related to the second $\phi$-derivative of $\ln V$, explicitly so in the $\varepsilon_{\phi\infty}\rightarrow 0$ limit.

In Friedmann equations, the matter density term is proportional to $(1+z)^3$ and the spatial curvature term to $(1+z)^2$. Therefore, a small $|\Omega_k|\lesssim \Omega_m$ has negligible impact at high redshift. We thus expect most of the $\Omega_k$ effect to be captured by replacing $F(0, x)$ in HBK's three-parameter parametrization with $F(\lambda, x)$. On the other hand, in the limit where $\frac{\epsilon_V\rho_\phi}{3H^2M_p^2}$ is exactly a constant, the ``slow-roll'' terms, which alter the tracking solution at low redshift, should exactly vanish. This requirement suggests a replacement:
\begin{equation}
  \sqrt{\varepsilon_s} - \sqrt{2\varepsilon_{\phi\infty}} \rightarrow \sqrt{\varepsilon_s} - \sqrt{\frac{2\varepsilon_{\phi\infty}}{1-\Omega_k}}.
\end{equation}
Finally we arrive at a generalized 3-parameter parametrization:
\begin{equation}
  w = -1 + \frac{2}{3}\left\{\sqrt{\varepsilon_{\phi\infty}} + \left(\sqrt{\varepsilon_s} - \sqrt{\frac{2\varepsilon_{\phi\infty}}{1-\Omega_k}}\right)\left[F\left(\frac{\Omega_ka_{\rm eq}}{\Omega_m}, \frac{a}{a_{\rm eq}}\right) + \zeta_s F_2\left(\frac{a}{a_{\rm eq}}\right)\right]\right\}^2\,, \label{eq:3param}
\end{equation}
where $a_{\rm eq}$ can be approximated by
\begin{equation}
  a_{\rm eq} \equiv \left(\frac{\Omega_m}{\Omega_\phi}\right)^{\frac{1}{3-\delta}}, \label{eq:aeq}
\end{equation}
and
\begin{eqnarray}
  \delta \equiv && \left\{\sqrt{\varepsilon_{\phi\infty}}+ \left[0.91-\frac{0.78\Omega_m}{1-\Omega_k}+ (0.24-\frac{0.76\Omega_m}{1-\Omega_k})\zeta_s\right] \left(\sqrt{\varepsilon_s}-\sqrt{\frac{2\varepsilon_{\phi\infty}}{1-\Omega_k}}\right)\right\}^2  \nonumber \\
  && + \left[\sqrt{\varepsilon_{\phi\infty}}+\left(0.53-0.09\zeta_s\right)\left(\sqrt{\varepsilon_s}-\sqrt{\frac{2\varepsilon_{\phi\infty}}{1-\Omega_k}}\right)\right]^2  \label{eq:delta} 
\end{eqnarray}
is a fitting formula describing the variation of dark energy density in the slow-roll regime ($a\ge a_{\rm eq}$). The ``shape correction'' function
\begin{equation}
F_2 (x) \equiv \sqrt{2}\left[1-{\ln \left(1+x^3\right) \over x^3}\right] - \frac{\sqrt{1+x^3}}{x^{3/2}}+\frac{\ln\left[x^{3/2}+\sqrt{1+x^3}\right]}{x^3} \,  \label{eq:F2def}
\end{equation}
formulates the impact of the ``running parameter'' $\zeta_s$.

\begin{figure*}
  \includegraphics[width=0.333\textwidth]{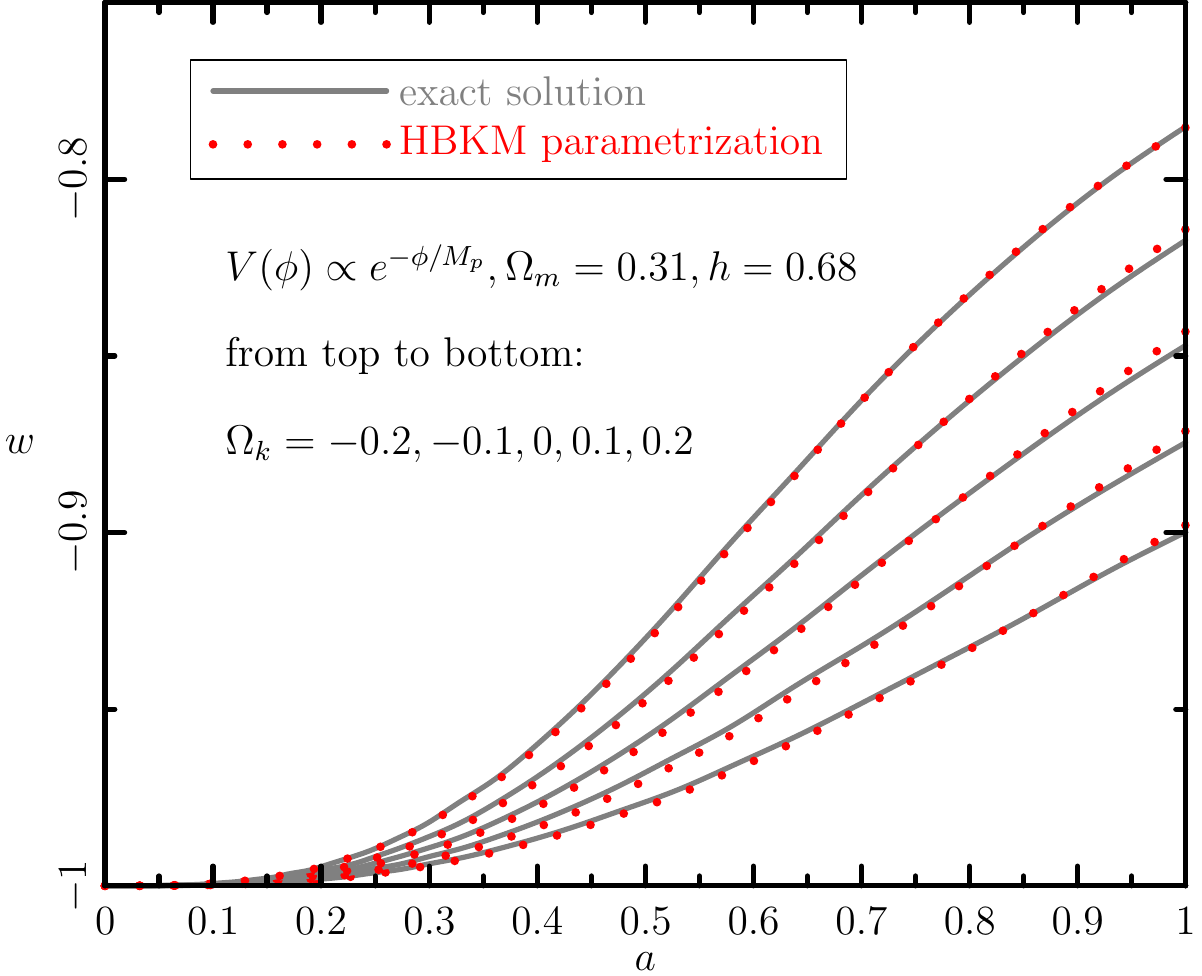} %
  \includegraphics[width=0.333\textwidth]{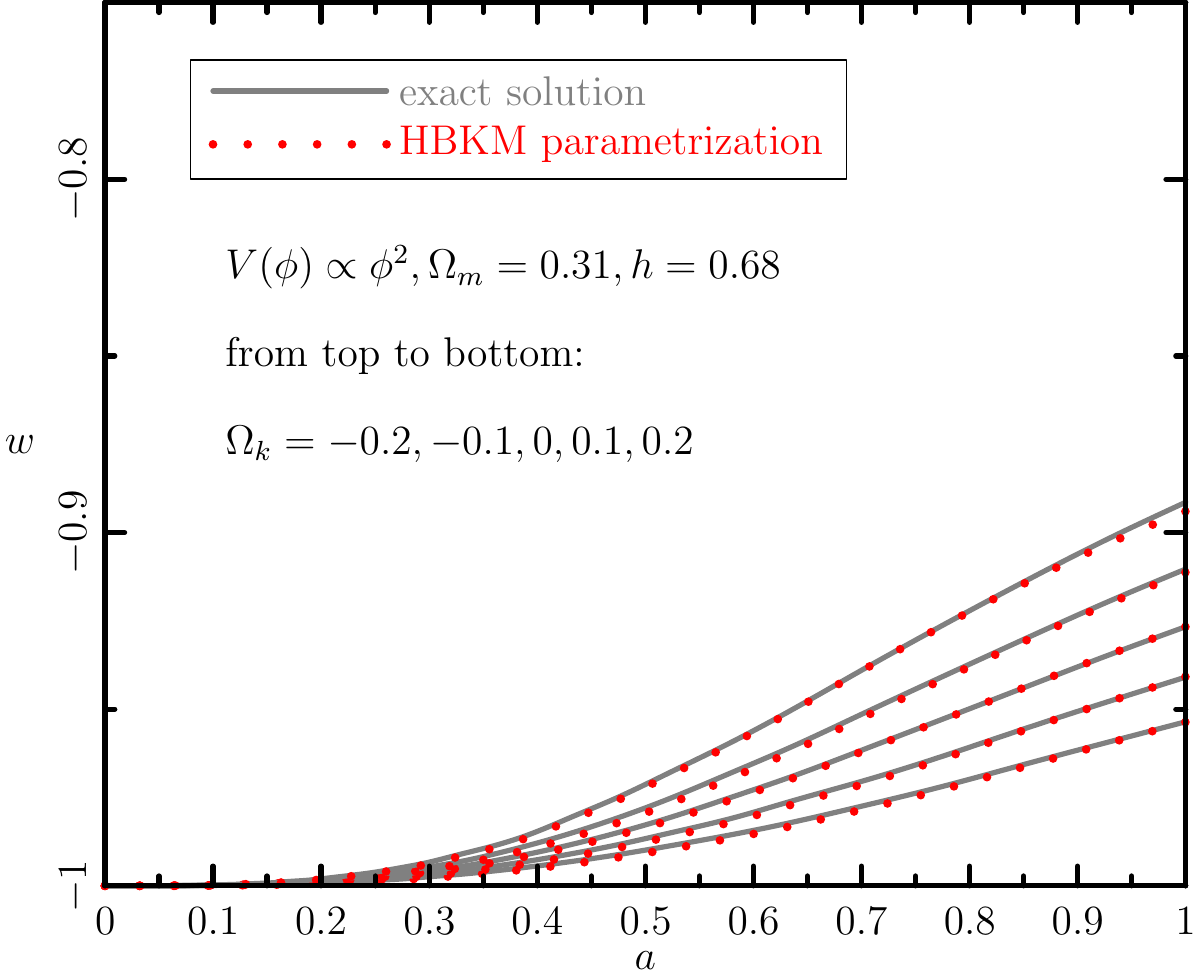} %
  \includegraphics[width=0.333\textwidth]{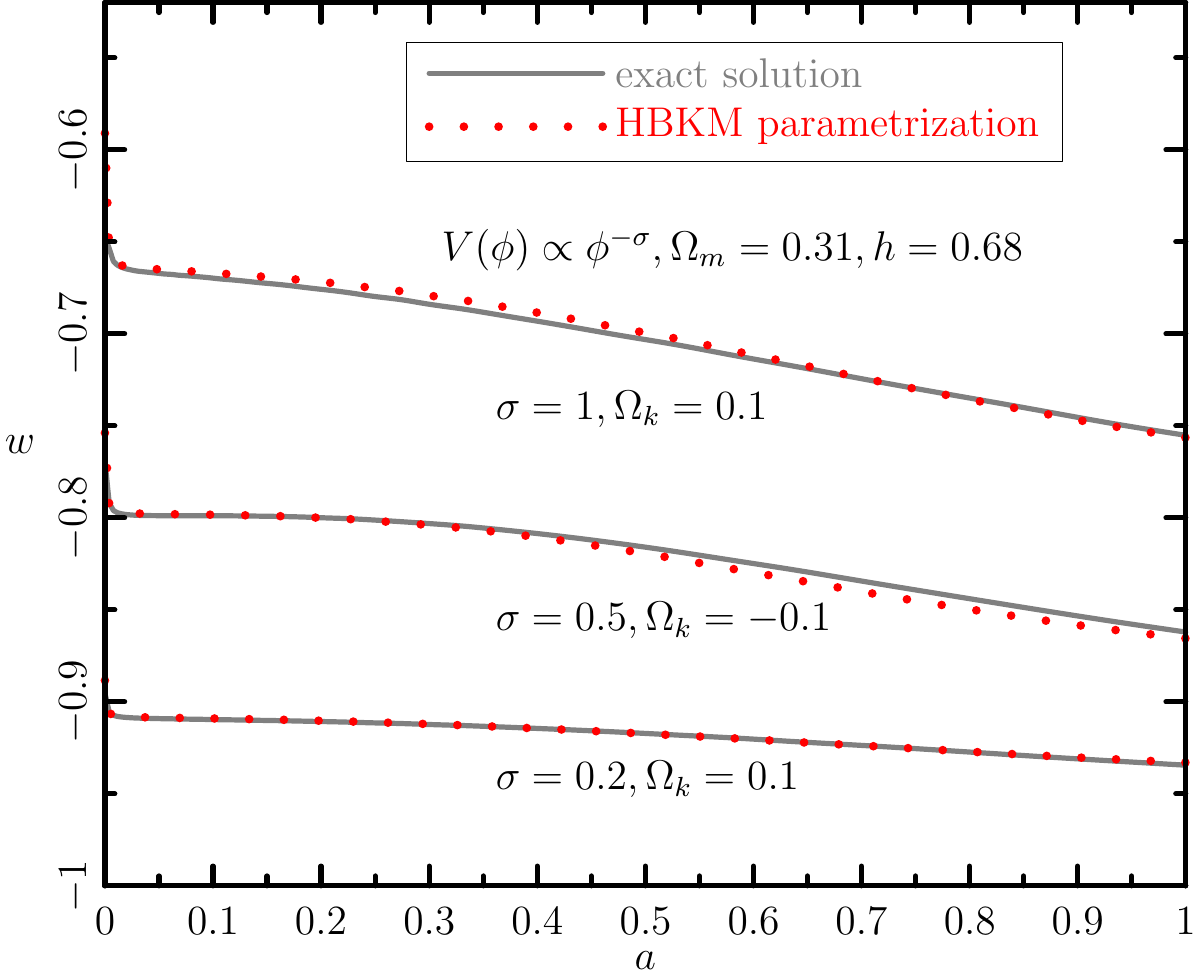}   
  \caption{Examples of $w(a)$ trajectories for exponential potential (left panel), power-law potential (middle panel) and negative power-law potential (right panel). For the flat-potential models (left and middle panels) the initial condition is given by $\dot\phi|_{a\rightarrow 0^+} \rightarrow 0$, whereas for the tracking models (right panel) the initial condition is given by its tracking solution. Solid gray lines are the exact $w(a)$ solutions. Dotted red lines are HBKM parametrization, i.e., Eq.~\eqref{eq:3param}. \label{fig:examples}}
\end{figure*}

A few concrete examples are given in Figure~\ref{fig:examples} to illustrate the accuracy of HBKM (Huang, Bond, Kofman, Miao) parametrization, i.e., the generalized HBK parametrization in Eq.~\eqref{eq:3param}.

\section{Data sets}\label{sec:data}

To compare parametrization~\eqref{eq:w} and \eqref{eq:3param} with observations, we use the publicly available software CosmoMC \citep{cosmomc} and replace its default CPL parametrization with Eq.~\eqref{eq:w} and Eq.~\eqref{eq:3param}. Our parametrization include $\varepsilon_s$, $\varepsilon_{\phi\infty}$, $\zeta_s$, $\Omega_k$, and the standard six parameters: the baryon density $\Omega_bh^2$, the cold dark matter density $\Omega_ch^2$, the angular extension of sound horizon on the last scattering surface $\theta_{MC}$, the CMB optical depth $\tau$, the primordial scalar metric fluctuation amplitude $A_s$ and its spectral index $n_s$. The Hubble constant $H_0$ and present matter density fraction $\Omega_m$ can be derived from these parameters. Flat priors $0\le \varepsilon_s\le 1.5$, $0\le \varepsilon_{\phi\infty}\le 1$, and $-1\le \zeta_s\le 1$ are used.

The following data sets are used for Monte Carlo Markov Chain (MCMC) calculations.

\subsection{Type Ia Supernovae}

Type Ia supernovae are known as standard candles at cosmological distance. The ``Joint Light-curve Analysis'' (JLA) samples, used in \cite{PlanckParam15} and this paper, is a joined data set of the Supernova Legacy Survey (SNLS) data \citep{SNLS} , Sloan Digital Sky Survey (SDSS) SNe data  \citep{SDSS-SNe}, and some low redshift SNe samples.

\subsection{Cosmic Microwave Background}

CMB is a powerful tool to measure the primordial fluctuations, the matter content, and the geometry of the universe. We use the full Planck 2015 release (TT,TE, EE + lowP + lensing). The details of Planck mission and its data description can be find in \cite{PlanckOverview15}, \cite{PlanckParam15}, \cite{PlanckLike15}, and references therein.

\subsection{Baryon Acoustic Oscillations}

BAO is a ``standard ruler'' that measures the geometry of the late-time universe. It is based on known physics and very few assumptions about the Universe, and is considered to be very reliable and almost free of systematics. Here we use the recent SDSS data release 12 \citep{BAO-SDSS-DR12-CMASS, BAO-SDSS-DR12-LOWZ} together with some low-redshift data sets \citep{BAO-SDSS-6DF, BAO-SDSS-DR7-MGS}. The full data set covers redshift up to $z\sim 0.6$.

\section{Results \label{sec:results}}

To demonstrate the impact of different data sets, we do MCMC calculation for CMB only, SNe + BAO, and all the three together, respectively, for the one-parameter parametrization Eq.~\eqref{eq:w}. With CMB only or SNe + BAO, the degeneracy between geometrical parameters are very strong. We use a very weak Gaussian prior $H_0 = 70.6 \pm 3.3 $ (see \cite{HST70p6}) to avoid the MCMC chains exploring nonphysical regions. With all the three data sets together (CMB + SNe + BAO), the degeneracy is not strong and we do not use the $H_0$ prior.

In Table~\ref{tbl:cosmoparams} we show the results of the abovementioned three runs for the  one-parameter parametrization and of a CMB + SNe + BAO run for the three-parameter parametrization Eq.~\eqref{eq:3param}.  The combination of SNe + BAO + weak $H_0$ prior only constrains background geometric at low redshift. Thus, in this case $\Omega_bh^2$ is perfectly degenerate with $\Omega_ch^2$ and only their sum $\Omega_mh^2$ can be constrained. 

\begin{table}
\caption{The median value and 68.3\% confidence level (CL) upper/lower limits of cosmological parameters. For $\varepsilon_s$ and $\varepsilon_{\phi\infty}$ that are bounded from below by the theory, 68.3\%CL upper-limit and 95.4\% upper-limit are shown. A dash indicates unused parameter, while ``unconstrained'' means that the parameter is not constrained by the data (posterior $\approx$ prior).}
\label{tbl:cosmoparams}
  \begin{center}
\begin{tabular}{l|l|lll}
\hline
\hline
& 3-parameter & & 1 - parameter  & \\
& CMB + SNe + BAO & CMB + SNe + BAO & CMB + weak $H_0$ prior & SNe + BAO + weak $H_0$ prior \\
\hline
$\Omega_b h^2$ & $0.02223^{+0.00017}_{-0.00017}$ & $0.02223^{+0.00016}_{-0.00016}$  & $0.02224^{+0.00016}_{-0.00017}$ & unconstrained \\
\hline
$\Omega_c h^2$ & $0.1194^{+0.0016}_{-0.0015}$ &  $0.1194^{+0.0015}_{-0.0015}$ & $0.1195^{+0.0015}_{-0.0015}$ & unconstrained  \\
\hline
$100\theta_{MC}$ & $1.04082^{+0.00033}_{-0.00034}$ &  $1.04084^{+0.00036}_{-0.00034}$ & $1.04082^{+0.00035}_{-0.00034}$  & $1.25^{+0.10}_{-0.12}$ \\
\hline
$\tau$ & $0.073^{+0.014}_{-0.013}$ & $0.067^{+0.013}_{-0.013}$ & $0.069^{+0.016}_{-0.016}$ & - \\
\hline
$\Omega_K$ & $0.0036^{+0.0028}_{-0.0024}$ & $0.0022^{+0.0022}_{-0.0021}$ & $0.0030^{+0.0053}_{-0.0058}$ & $-0.101^{+0.060}_{-0.046}$  \\
\hline
$\varepsilon_s$ & $0.00^{+0.16+0.36}$ & $0.00^{+0.29+0.56}$ & $0.00^{+0.55+1.12}$ & $0.00^{+0.37+0.63}$  \\
\hline
$\epsilon_{\phi\infty}$ & $0.00^{+0.28+0.67}$ & - & - & - \\ 
\hline
$\zeta_s$ & unconstrained & - & - & - \\ 
\hline
${\rm{ln}}(10^{10} A_s)$ & $3.079^{+0.027}_{-0.025}$ & $3.067^{+0.024}_{-0.023}$& $3.07^{+0.03}_{-0.03}$ & -\\
\hline
$n_s$ & $0.965^{+0.005}_{-0.005}$ & $0.965^{+0.005}_{-0.005}$  & $0.965^{+0.005}_{-0.005}$ & - \\
\hline
$H_0$ & $67.3^{+0.8}_{-0.9}$ & $67.3^{+0.8}_{-1.1}$ & $66.7^{+2.7}_{-2.5}$  & $69.9^{+2.6}_{-3.0}$ \\
\hline
$\Omega_m$ & $0.314^{+0.009}_{-0.008}$ & $0.314^{+0.009}_{-0.008}$  & $0.321^{+0.025}_{-0.024}$ &  $0.340^{+0.028}_{-0.032}$ \\
\hline
\end{tabular}
\end{center}
\end{table}

\begin{figure}
  \centering
  \includegraphics[width=0.48\textwidth]{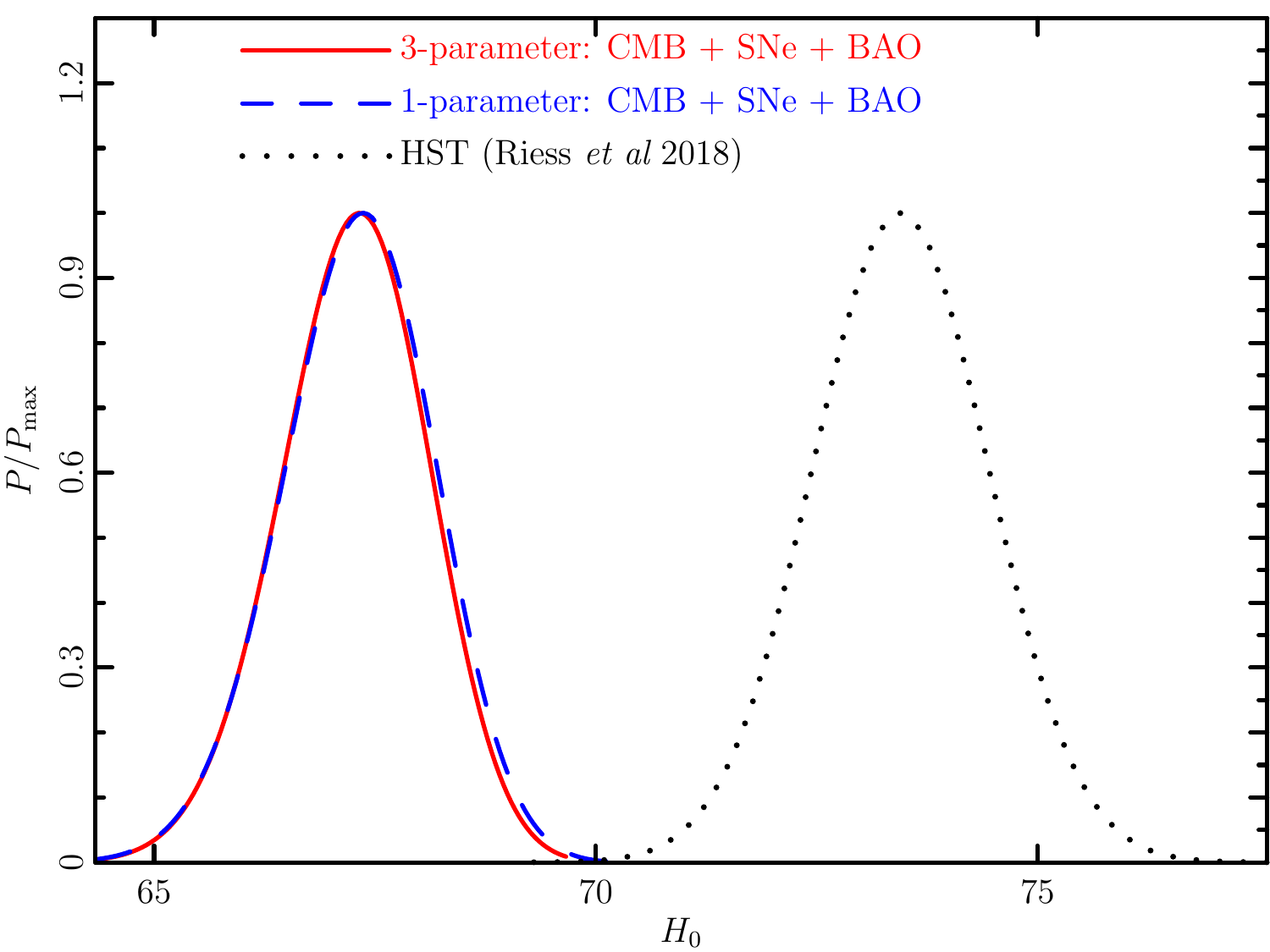}
  \caption{ Marginalized constraint on $H_0$ with CMB + SNe + BAO compared with the local $H_0$ measurement from \cite{Riess18}. \label{fig:H0}}
\end{figure}

In the three-parameter parametrization case with CMB + SNe + BAO, the posterior of $H_0$ is close to a Gaussian distribution $H_0 = 67.25\pm 0.84\, \mathrm{km/s/Mpc}$, which is in $3.3\sigma$ tension with $H_0 = 73.48 \pm 1.66\, \mathrm{km/s/Mpc}$ from \cite{Riess18}. In the one-parameter case, the deviation of $H_0$ posterior from a Gaussian distribution ($67.30\pm 0.86$ constructed from mean and standard deviation) is more significant. While using the full posterior distribution constructed from the MCMC calculation, we find the tension slightly drops to $3.2\sigma$, which is not a significant change of the story. In the left panel of Figure~\ref{fig:H0}, we visualize the $H_0$ tension by plotting the CMB + SNe + BAO posterior, for both one-parameter and three-parameter parametrizations, against the \cite{Riess18} result.

\begin{figure*}
  \includegraphics[width=0.48\textwidth]{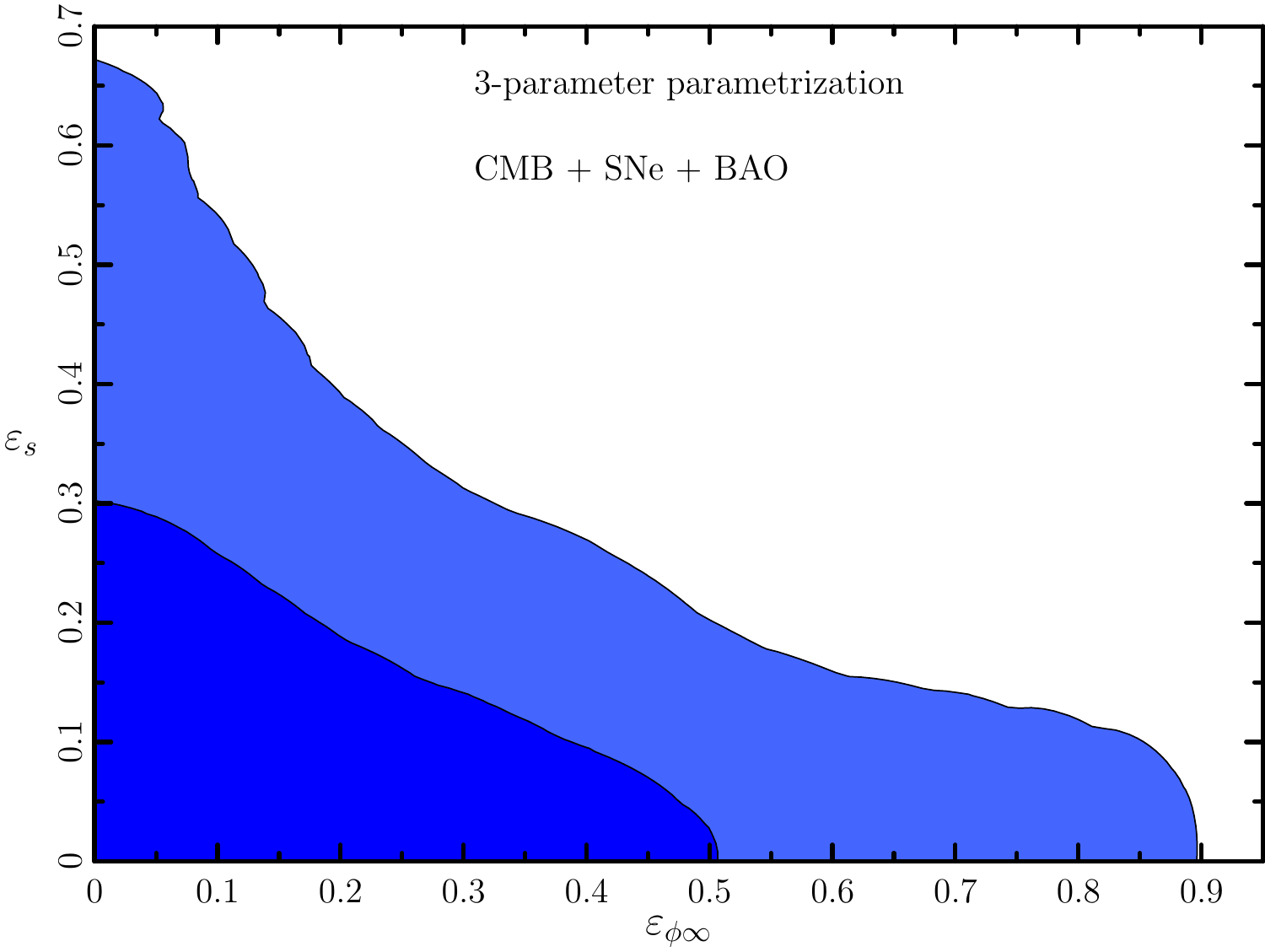}\hspace{0.02\textwidth}%
  \includegraphics[width=0.48\textwidth]{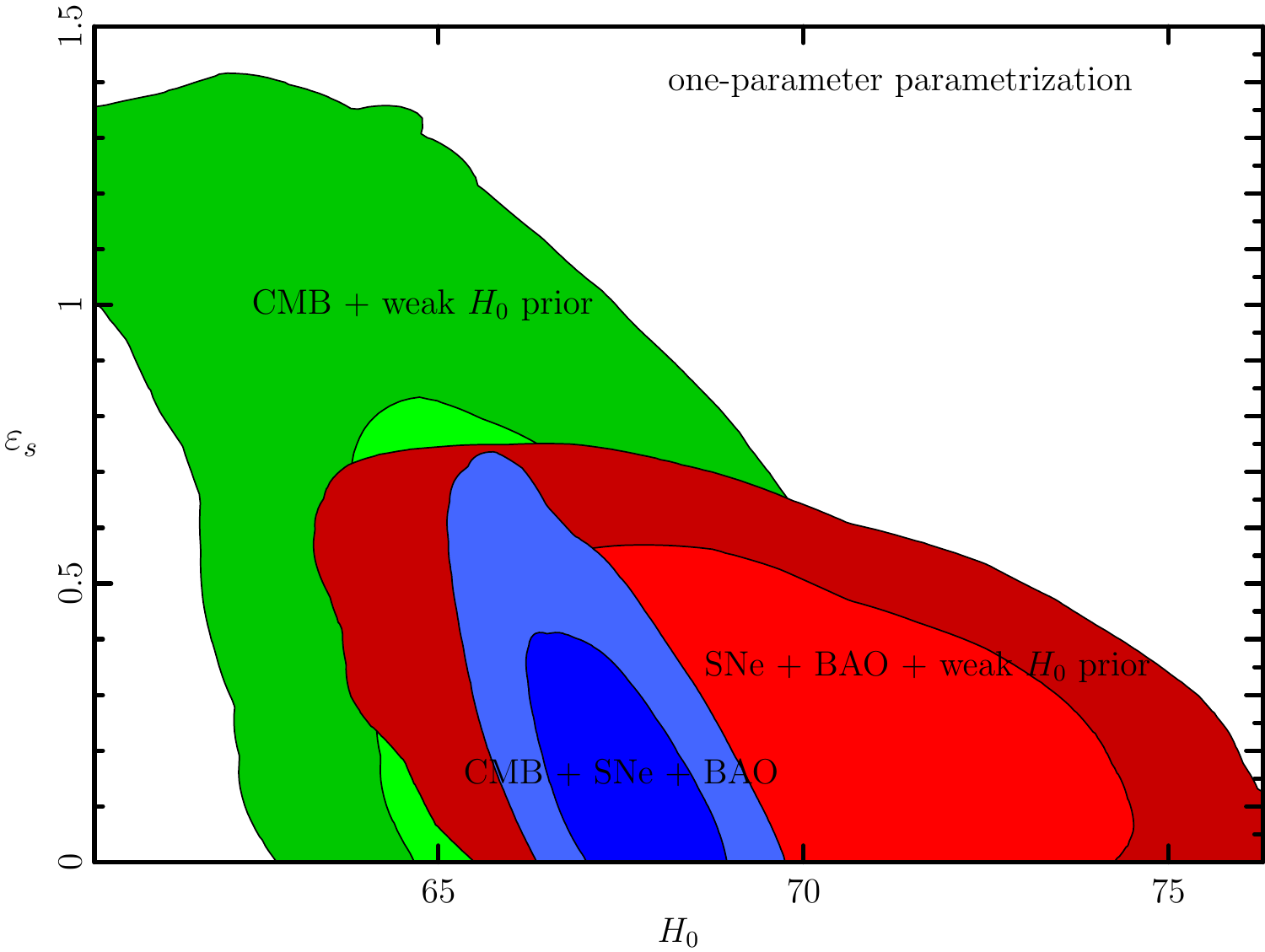}  
  \caption{Left panel: marginalized 68.3\% CL (inner contour) and 95.4\% CL (outer contour) constraints on $\varepsilon_s$ and $\varepsilon_{\phi\infty}$ (three-parameter parametrization); right panel: marginalized 68.3\% CL (inner contours) and 95.4\% CL (outer contours) constraints on $H_0$ and $\varepsilon_s$ (one-parameter parametrization) with different combinations of data sets. \label{fig:2d}}
\end{figure*}

In three-parameter parametrization  $\varepsilon_s$ is ``anomalously'' constrained better than in one-parameter case. Similar results have been shown in \cite{HBK} and \cite{PlanckDE15},  and were understood as an effect due to the anti-correlation between $\varepsilon_s$ and $\varepsilon_{\phi\infty}$ shown in the left panel of Figure~\ref{fig:2d}.

In the right panel of Figure~\ref{fig:2d} we show the impact of different data sets on the constraints on $\varepsilon_s$ and $H_0$ for one-parameter parametrization. We find $\varepsilon_s$ and $H_0$ slightly anti-correlated. Thus, compared to $\Lambda$CDM $\varepsilon_s=0$ case, the relaxation of $\varepsilon_s$ may even worsen the $H_0$ tension.

Although the combined low-redshift data SNe + BAO + weak $H_0$ prior slightly prefer a negative $\Omega_k$ ($\sim 1.7\sigma$ level),  the CMB data drive $\Omega_k$ back toward zero and give much tighter bounds. We thus have justified the flatness assumption used in \cite{HBK} with a fully self-consistent calculation.

\section{Conclusion and Discussion \label{sec:conclusion}}

The $H_0$ tension between \cite{PlanckParam15} (assuming $\Lambda$CDM) and \cite{Riess18}, if taken at face value, suggests evidence for new physics beyond $\Lambda$CDM at more than $99\%$ confidence level. We explored one of the simplest alternatives to $\Lambda$CDM: a slowly-or-moderately rolling quintessence field with a smooth potential in a non-flat FRW universe. We generalized the model-independent HBK parametrization to non-flat FRW metric. Using the latest CMB, SNe and BAO data, we find that the four additional degrees of freedom ($\varepsilon_s$, $\varepsilon_{\phi\infty}$, $\zeta_s$ and the spatial curvature $\Omega_k$) do not ease, if not worsen, the tension between local and high-redshift measurements of the Hubble constant.

There are many other interesting attempts to explore the $H_0$ tension with alternative cosmologies. Among those some are phenomenological models, such as free lensing amplitude~\citep{Grandis16}, late-time spatial curvature~\citep{Bolejko18}, \"u$\Lambda$CDM model~\citep{Khosravi17}, and XCDM cosmology~\citep{Ooba18} , and the others are self-consistent models, such as Galileon gravity~\citep{Renk17} and hot axions~\citep{HotAxions}. Some phenomenological models, despite their lack of physical consistency, can mitigate the $H_0$ tension. 

The $H_0$ tension may also be subject to some unknown observational biases and systematics. \cite{HST70p6} argued that there might be uncounted systematic effects in HST Cepheid calibration. Thanks to the recent Gaia data release 2 \citep{GaiaDR2}, \cite{H0GaiaDR2} was able to do a more accurate calibration and found no significant migration of Cepheid standards. Meanwhile, many other efforts have been made to obtain an unbiased $H_0$ from existing $H_0$ and $H(z)$ measurements. \cite{Chen11} applied median statistics, which is supposed less sensitive to outliers with unknown systematics, on 553 $H_0$ measurements and obtained $68\pm 5.5\,\mathrm{km\,s^{-1}\,Mpc^{-1}}$.  More recently, \cite{Yu18} used Gaussian Process method to determine a continuous $H(z)$ function, and found  a similar result $H_0 = 67\pm 4\,\mathrm{km\,s^{-1}\,Mpc^{-1}}$. This result was soon updated to $67.06\pm 1.68\,\mathrm{km\,s^{-1}\,Mpc^{-1}}$ by \cite{Gomez18}, who added Type Ia supernovae into their analysis. These measurements, independent of HST and Planck constraints, seem to favor a lower $H_0$ value that is more consistent with Planck result. This conclusion was further embraced by \cite{Zhang18}, who combined BAO measurement with tomography Alcock-Paczinsky method, and \cite{Addison18}, who did a more detailed study by combining galaxy BAO data with a variety of data that are independent of HST and Planck measurements.

\end{document}